\documentclass[usenatbib]{mn2e}
\usepackage{graphicx}
\usepackage{deluxetable}

\def\apj{\rm ApJ}
\def\apjl{\rm ApJL}
\def\apjs{\rm ApJS}

\def\aap{\rm AAP}
\def\araa{\rm ARA\&A}

\def\gax{\mathrel{\raise.3ex\hbox{$>$}\mkern-14mu\lower0.6ex\hbox{$\sim$}}}
\def\lax{\mathrel{\raise.3ex\hbox{$<$}\mkern-14mu\lower0.6ex\hbox{$\sim$}}}
\def\gtorder{\mathrel{\raise.3ex\hbox{$>$}\mkern-14mu
             \lower0.6ex\hbox{$\sim$}}}
\def\ltorder{\mathrel{\raise.3ex\hbox{$<$}\mkern-14mu
             \lower0.6ex\hbox{$\sim$}}}

\begin{document}

\title[Stellar Mergers Are Common]
   {Stellar Mergers are Common}

\author[C.~S. Kochanek et al.]{ C.~S. Kochanek$^{1,2}$, Scott M. Adams$^{1,2}$  and Krzysztof Belczynski$^{3,4}$\\
  $^{1}$ Department of Astronomy, The Ohio State University, 140 West 18th Avenue, Columbus OH 43210 \\
  $^{2}$ Center for Cosmology and AstroParticle Physics, The Ohio State University,
    191 W. Woodruff Avenue, Columbus OH 43210 \\
  $^{3}$ Astronomical Observatory, Warsaw University, Al.
    Ujazdowskie 4, 00-478 Warsaw, Poland (kbelczyn@astrouw.edu.pl)\\
  $^{4}$ Center for Gravitational Wave Astronomy, University of Texas at
    Brownsville, Brownsville, TX 78520
   }

\maketitle

\begin{abstract}
The observed Galactic rate of stellar mergers or the initiation of common envelope phases
brighter than $M_V=-3$ ($M_I=-4$) is of order
$\sim 0.5$ ($0.3$)~year$^{-1}$ with 90\% confidence statistical uncertainties of $0.24$--$1.1$
($0.14$--$0.65$) and factor of $\sim 2$ systematic uncertainties.  The (peak) luminosity function is 
roughly $dN/d L \propto L^{-1.4\pm0.3}$, so the rates for events more luminous than V1309~Sco 
($M_V\simeq -7$~mag) or V838~Mon ($M_V \simeq -10$~mag) are lower at $r\sim 0.1$/year
and $\sim 0.03$/year, respectively.  The peak luminosity is a steep function
of progenitor mass, $L \propto M^{2-3}$.  This very roughly parallels the scaling
of luminosity with mass on the main sequence, but the transients are $\sim 2000$-$4000$ times 
more luminous at peak.
Combining these, the mass function of the progenitors, $dN/dM \propto M^{-2.0\pm0.8}$, is
consistent with the initial mass function, albeit with broad uncertainties.  
These observational results are also broadly consistent with the estimates of 
binary population synthesis models.
While extragalactic variability surveys can better define the rates and properties of the
high luminosity events, systematic, moderate depth ($I\gtorder 16$~mag) surveys of the Galactic plane are needed
to characterize the low luminosity events.  The existing Galactic samples are only $\sim 20\%$
complete and Galactic surveys are (at best!) reaching a typical magnitude limit of $\ltorder 13$~mag.   
\end{abstract}


\section{Introduction}
\label{sec:introduction}

In January 2002, the transient V838~Mon was discovered (\citealt{Brown2002})
and then produced a series of dust
echoes in the surrounding interstellar medium (ISM) that are some of the most iconic
images produced by the {\it Hubble Space Telescope} (HST, \citealt{Bond2003}).  In September 2008, the 
transient V1309~Sco was discovered (\citealt{Nakano2008}), with spectacular evidence from OGLE light curves
that the source was a binary merger (\citealt{Tylenda2011}).  
V4332~Sgr in February 1994
and OGLE~2002-BLG-360 in October 2002 appear to be similar events
(\citealt{Martini1999}, \citealt{Tylenda2013}), and CK~Vul (Nova Vul 1670)
has been proposed as a historical example (\citealt{Kato2003}).
The M31~RV (``Red Variable'') and M85~OT2006-1 are possible extragalactic
examples (\citealt{Rich1989}, \citealt{Kulkarni2007}, but see
\citealt{Pastorello2007} and \citealt{Thompson2009} for alternative
interpretations of the M~85 transient). Other lower luminosity
extragalactic transients, such as SN~2008S and the 2008 transient
in NGC~300, almost certainly belong to a different class of objects
(see \citealt{Prieto2008}, \citealt{Thompson2009},
\citealt{Kochanek2011}).
 
While there has been some argument for interpreting these events as
an odd form of nova (e.g. \citealt{Shara2010}), the generally accepted interpretation
is that they are stellar mergers or the dynamical phase of common envelope (CE)
evolution as originally proposed
by \cite{Soker2003}.  The arguments in favor of gravitationally
driven mergers over thermonuclear novae are laid out by 
\cite{Tylenda2006}, but this issue became moot with the direct
observations of V1309~Sco as a merging binary (\citealt{Tylenda2011}).
Stellar mergers are relatively gentle processes (e.g. \citealt{Passy2012}, 
\citealt{Ricker2012}, \citealt{Nandez2013}), producing
only limited mass loss and leaving a remnant with a vastly inflated
envelope appearing as an M (or even L) type supergiant combined
with dust formation (as first
noted by \cite{Rich1989} for the M31~RV and \cite{Martini1999}
for V4332~Sgr).  
In particular, the survival of the
circumstellar dust surrounding OGLE~2002-BLG-360 (\citealt{Tylenda2013})
means that no fast, radiative shock was associated with the transient.
For the SN~2008S class of transients (\citealt{Thompson2009}), almost
all the circumstellar dust must be destroyed in the transient which
requires an unobserved luminosity spike ($\sim 10^{10}L_\odot$!) that can 
only be explained by a fast
shock breaking out from the surface of the star 
(see \citealt{Kochanek2011}).

These events are not curiosities, but a powerful probe of
binary evolution.  Most stars are in 
binaries, and a non-trivial fraction are in triples or 
higher order systems (see, e.g., the review by \citealt{Duchene2013}). 
As binary stars evolve, they frequently interact 
(e.g., recently, \citealt{Sana2012} and \citealt{deMink2014}
for massive stars)
or are driven to interact by a tertiary 
(e.g. \citealt{Fabrycky2007},
\citealt{Thompson2011}) leading to 
a CE phase and potentially a full merger.  To somewhat simplify 
our language we will refer to both of these as mergers, although the ``cores'' 
of the stars need not ultimately merge.  These interactions are
not only theoretically expected, but are also required to explain
many classes of compact binaries or peculiar stars such
as blue stragglers (e.g., \cite{Bailyn1995} for globular
clusters) and R~Cor~Bor stars (e.g. \citealt{Webbink1984}).
Recent studies of massive stars suggest that $\sim 10\%$ of massive 
stars are probably merger products (\citealt{Sana2012}, \citealt{deMink2014}).
Directly determining the rates of such events would be a powerful
new constraint on models.

Despite the importance of making such measurements, there has been little 
discussion of the merger rates implied
by the existing events.  \cite{Soker2006} argue from analogy to estimates
of blue straggler formation that the Galactic rate should
be once every $10$-$50$~years.  \cite{Ofek2008} mention in passing an 
estimated empirical lower limit of $0.019$~year$^{-1}$, which 
was used for an estimated discovery rate by the 
Palomar Transient Factory (PTF, \citealt{Rau2009}, \citealt{Law2010}).   
{\it However, if the four Galactic events have a common
origin, these estimates are low -- with four such events in 25 years the 
implied rate of Galactic stellar mergers is $>0.1$/year even 
before making any corrections for completeness.}   
 
While there will be systematic uncertainties, in \S\ref{sec:rates} we estimate 
the rates and luminosity function of these events based on the statistics of
the Galactic events and then estimate the statistics for external galaxies.
In \S\ref{sec:properties} we examine the scaling of the event luminosities
with estimates of the progenitor masses.  In \S\ref{sec:theory} we compare
these observational results to estimates based on a binary population
synthesis model.  In \S\ref{sec:discuss} we discuss 
the implications of these results, the need for better variability surveys 
of the Galaxy, and the detectability of these events in other nearby galaxies.

\section{The Rate and Luminosity Function}
\label{sec:rates}

Empirically, there have been four merger events in our galaxy over the last
$\Delta t = 25$ years, so the 90\% confidence limit on the rate $r$ is 
$ 2.0 \ltorder r C \Delta t \ltorder 9.2$  where $ C\leq 1$ is the
completeness.  This corresponds to having events every $r^{-1}=2.7$ to
$12.6 (25/\Delta t) C^{-1}$~years.  While there will be many
systematic uncertainties, we can estimate the completeness
and the luminosity function of these transients.  

In the Appendix we sketch the properties of the individual transients and
summarize their properties in Table~\ref{tab:summary}.  Almost all entries
in Table~\ref{tab:summary} other than peak magnitudes are approximate, but
for our present purposes the uncertainties in distances, extinctions and masses will
be unimportant.  We model the rate as a power law
\begin{equation}
     { d N \over dt d L } = A { r \over L_0} \left( { L \over L_0 }\right)^{-x}
\end{equation}
for $L_0 < L < L_1$, where $r$ is the global rate and
$A=(x-1)/(1-(L_1/L_0)^{1-x})$ is a normalization constant.  We assume a ``survey''
duration of $\Delta t = 25$~years which includes all the events from
V4332~Sgr onwards.
The completeness depends
on the fraction $f_{sky}$ of the sky (or Galaxy) that is surveyed and
the limiting magnitude $m_{lim}$ of the surveys.  For simplicity we
will model this as an all-sky/Galaxy survey, so $f_{sky} \equiv 1$,
and explore the scalings with $m_{lim}$ below.  All the rate
estimates below scale as $ r \propto f_{sky}^{-1} \Delta t^{-1}$.

To estimate the completeness we need $f(M)$, the probability that a
Galactic transient peaking at absolute magnitude $M$ will be discovered
given the effects of distance and extinction (but not sky coverage).
In \cite{Adams2013} we made such estimates for the visibility of
Galactic supernovae, and we simply adopt their models. Core
collapse supernovae (ccSNe) were assumed to be distributed
following a standard Galactic thin disk model (TRILEGAL,
\citealt{Girardi2005}) with a scale height of $H=95$~pc, while thermonuclear Type~Ia SN were
half distributed in the thin disk (the ``prompt component'')
and half in the thick disk (the ``delayed component'') 
with a scale height of $H=800$~pc.
The extinction along any site line was normalized to an
empirical estimate of the extinction and then distributed
to follow the thin disk density in that direction.
Compared to other systematic problems, uncertainties in
distances and the effects of inhomogeneities such as spiral arms
are unimportant.  We will refer to these two spatial
distributions as the thin disk and the thick$+$thin disk
distributions, where the practical difference is that the
thick disk component suffers far less extinction than the
thin disk component. 

We only consider the ``standard'' extinction model from \cite{Adams2013}
using the Rayleigh-Jeans Color Excess (RJCE) extinction maps
from \cite{Nidever2012} in the Galactic plane and reverting to a modified
version (\citealt{Bonifacio2000}) of the \cite{Schlegel1998} maps where the
RJCE estimates are unavailable.
We considered several other extinction normalizations in \cite{Adams2013},
but the differences are unimportant to our completeness
corrections here because they only matter for
high ($E(B-V)>1$) extinctions.  All the observed transients
were very bright and only moderately extincted (see Table~\ref{tab:summary}),
so changes in the dust model normalizations have  negligible effects on the
rate estimate.  In \cite{Adams2013} we found this was also true
when estimating supernova rates based on the very bright
historical supernovae seen in the Galaxy.

The observed transients all lie in or near the Galactic disk
(see Table~\ref{tab:summary}), so we view
the thin disk model as more representative. We provide the results
for the thick$+$thin disk model primarily as a contrast.  More realistically, the
spatial distribution is mass-dependent.  The more massive progenitors
($M \gtorder 2M_\odot$) must all be associated with the thin disk
simply because they are young, while the lower mass
systems will have scale heights that increase with age, reaching
a maximum comparable to the thick disk component.  However,
the differences even between the results for the two spatial
distributions are not that large given the other uncertainties, so a more complex
model is presently unwarranted.

\onecolumn
\begin{deluxetable}{ccrccc}
\tablewidth{0pt}
\tablecaption{Observational Galactic Rate Estimates }
\tablehead{
  Model
 &Band 
 &$m_{lim}$ 
 &Rate $r$ 
 &Completeness $C$
 &Slope $x$ \\
  
 &
 &(mag)
 &(year$^{-1}$)
 &
 &
 } 
\startdata
thin &V & 9 &2.26 (0.67-7.00) &0.04 (0.02-0.12) &1.58 (1.13-2.10) \\
thin &V &11 &0.99 (0.36-2.40) &0.11 (0.07-0.19) &1.46 (1.15-1.79) \\
thin &V &13 &0.53 (0.22-1.10) &0.21 (0.15-0.30) &1.36 (1.10-1.66) \\
thin &I & 9 &1.86 (0.47-6.50) &0.05 (0.03-0.15) &1.72 (1.28-2.20) \\
thin &I &11 &0.66 (0.27-1.40) &0.17 (0.12-0.27) &1.53 (1.25-1.86) \\
thin &I &13 &0.32 (0.14-0.63) &0.35 (0.27-0.48) &1.43 (1.16-1.74) \\
thick$+$thin   &V & 9 &0.44 (0.19-0.92) &0.25 (0.18-0.38) &1.39 (1.12-1.70) \\
thick$+$thin   &V &11 &0.25 (0.11-0.52) &0.44 (0.36-0.56) &1.28 (1.01-1.58) \\
thick$+$thin   &V &13 &0.18 (0.06-0.47) &0.56 (0.48-0.69) &1.26 (0.89-1.63) \\
thick$+$thin   &I & 9 &0.33 (0.14-0.66) &0.34 (0.25-0.49) &1.45 (1.17-1.78) \\
thick$+$thin   &I &11 &0.17 (0.06-0.43) &0.58 (0.49-0.76) &1.35 (0.98-1.75) \\
thick$+$thin   &I &13 &0.13 (0.04-0.40) &0.73 (0.64-0.90) &1.33 (0.86-1.84) \\
\enddata
\label{tab:results}
\tablecomments{
The thin disk spatial distribution has sources only in a thin
disk with a scale height of $H=95$~pc, while the thick$+$thin
distribution puts half of the sources in a thick disk with a
scale height of $H=800$~pc (see \citealt{Adams2013}).
The estimates assume an all-sky search in the V or I bands to
a depth of $m_{lim}$ mag.  The median rate $r$, completeness $C$ and
luminosity function slope $x$ estimates are given along with their
90\% confidence ranges.  The rates can be further rescaled
as $r \propto f_{sky}^{-1} \Delta t^{-1}$ where we assumed
that fraction $f_{sky}=1$ of the Galaxy is surveyed for a period
of $\Delta t =25$~years.
}
\end{deluxetable}

\begin{deluxetable}{lcc}
\tablewidth{0pt}
\tablecaption{Theoretical Galactic CE Rates \hfill}
\tablehead{Donor/Companion & All CE & Mergers}
\startdata
MS/MS              & 0.088          & 0.088  \\
MS/Evolved         & 6.8 $10^{-6}$  & 6.8 $10^{-6}$ \\
MS/He star         & 2.4 $10^{-4}$  & 2.4 $10^{-4}$ \\
Ms/Compact         & 0.010          & 0.010 \\
                   &                &  \\
Evolved/MS         & 0.084          & 0.045 \\
Evolved/Evolved    & 2.4 $10^{-4}$  & 1.5 $10^{-4}$ \\
Evolved/He star    & 0.002          & 0.002 \\
Evolved/Compact    & 0.018          & 0.016 \\
                   &                &  \\
He star/any        & 0.001          & 2.2 $10^{-4}$ \\
                   &                &  \\
TOTAL              & 0.203          & 0.140 \\
\enddata
\label{tab:rates}
\tablecomments{
Rates are in units of year$^{-1}$ for either all CE events (middle)
or only the events predicted to end in a complete merger (right).
Stars off the main sequence are divided into evolved stars with
a hydrogen envelope and helium stars where the envelope has been
lost.  Compact objects include white dwarfs, neutron stars and
black holes.
}
\end{deluxetable}

\twocolumn

\begin{figure*}
\includegraphics[width=3.7in]{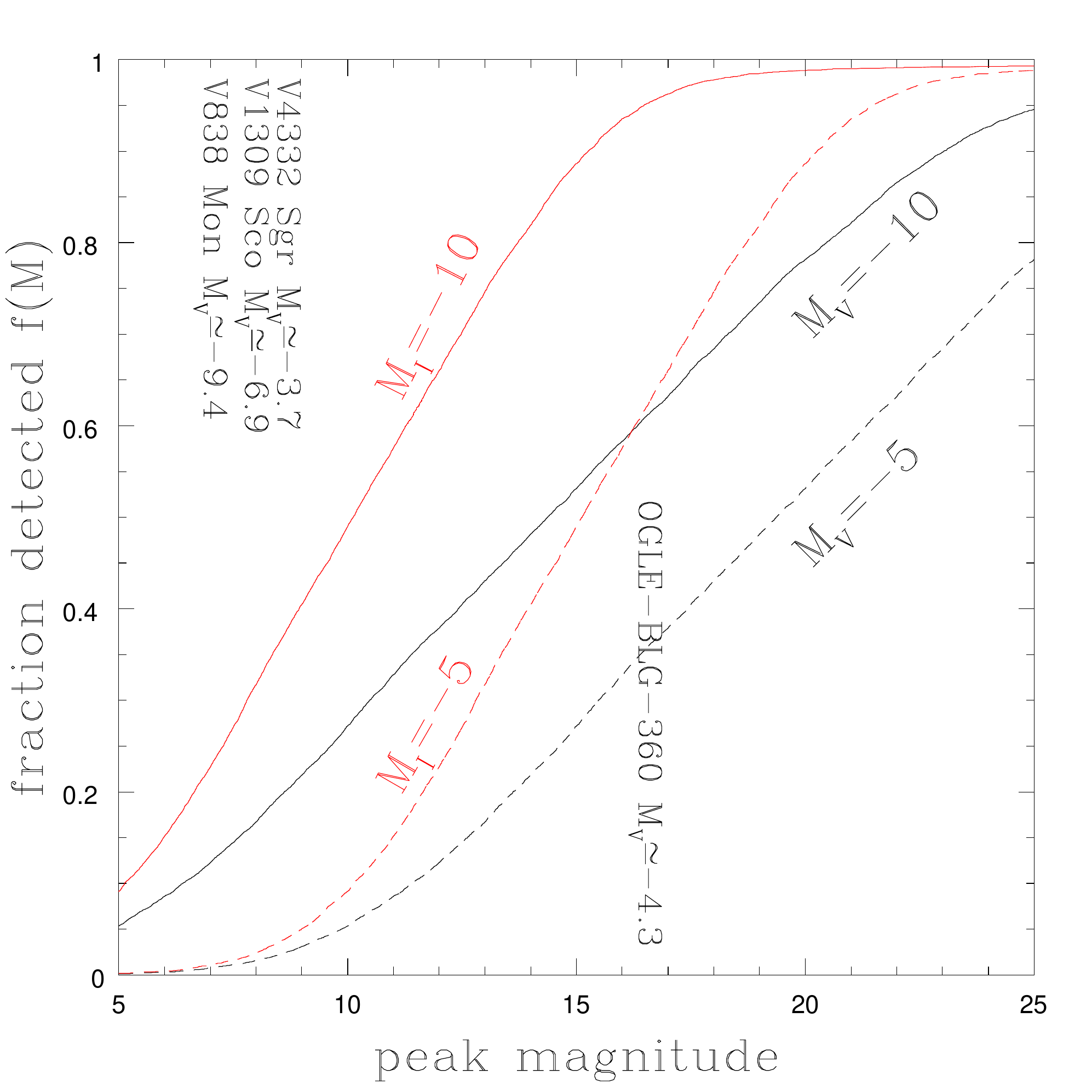}
\caption{
  Detection probabilities $f(M)$ for Galactic sources with 
  absolute V (black) or I (red) magnitudes of $-5$ (lower, dashed) 
  or $-10$ (upper, solid) as a function of apparent magnitude. 
  The peak V band magnitudes of the Galactic sources are marked.
  These estimates are for the thin disk spatial distribution.
  }
\label{fig:prob}
\end{figure*}

\begin{figure*}
\includegraphics[width=3.7in]{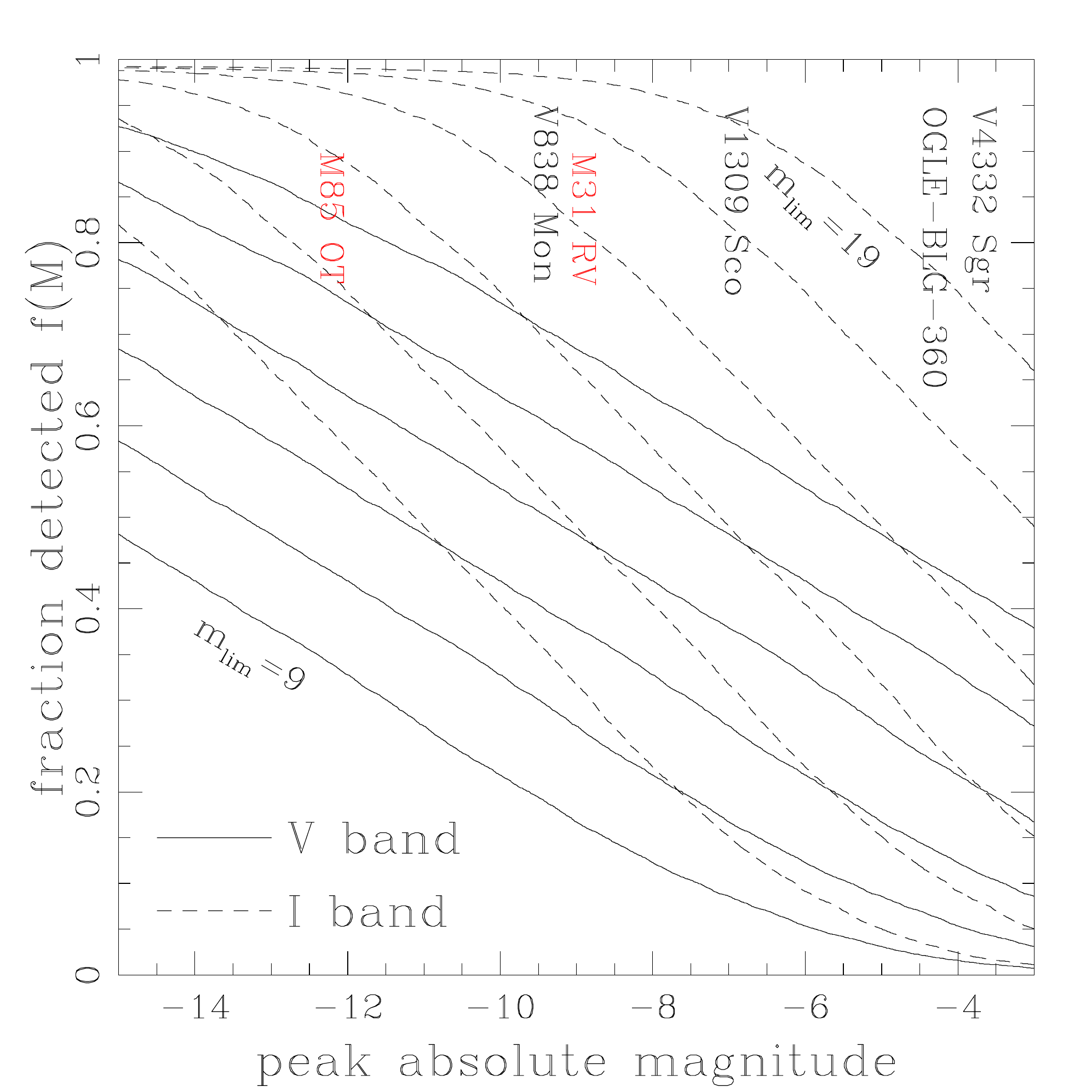}
\caption{
  Detection probabilities $f(M)$ for Galactic sources as
  a function of peak absolute V (solid) and I (dashed) magnitudes
  for limiting magnitudes of $m_{lim}=9$ (bottom), $11$, $13$,
  $15$, $17$ and $19$~mag (top).  
  The absolute V band magnitudes
  of the sources at peak are marked, including the two 
  extragalactic sources (red and lower).
  These estimates are for the thin disk spatial distribution.
  }
\label{fig:detect}
\end{figure*}

\begin{figure*}
\includegraphics[width=3.5in]{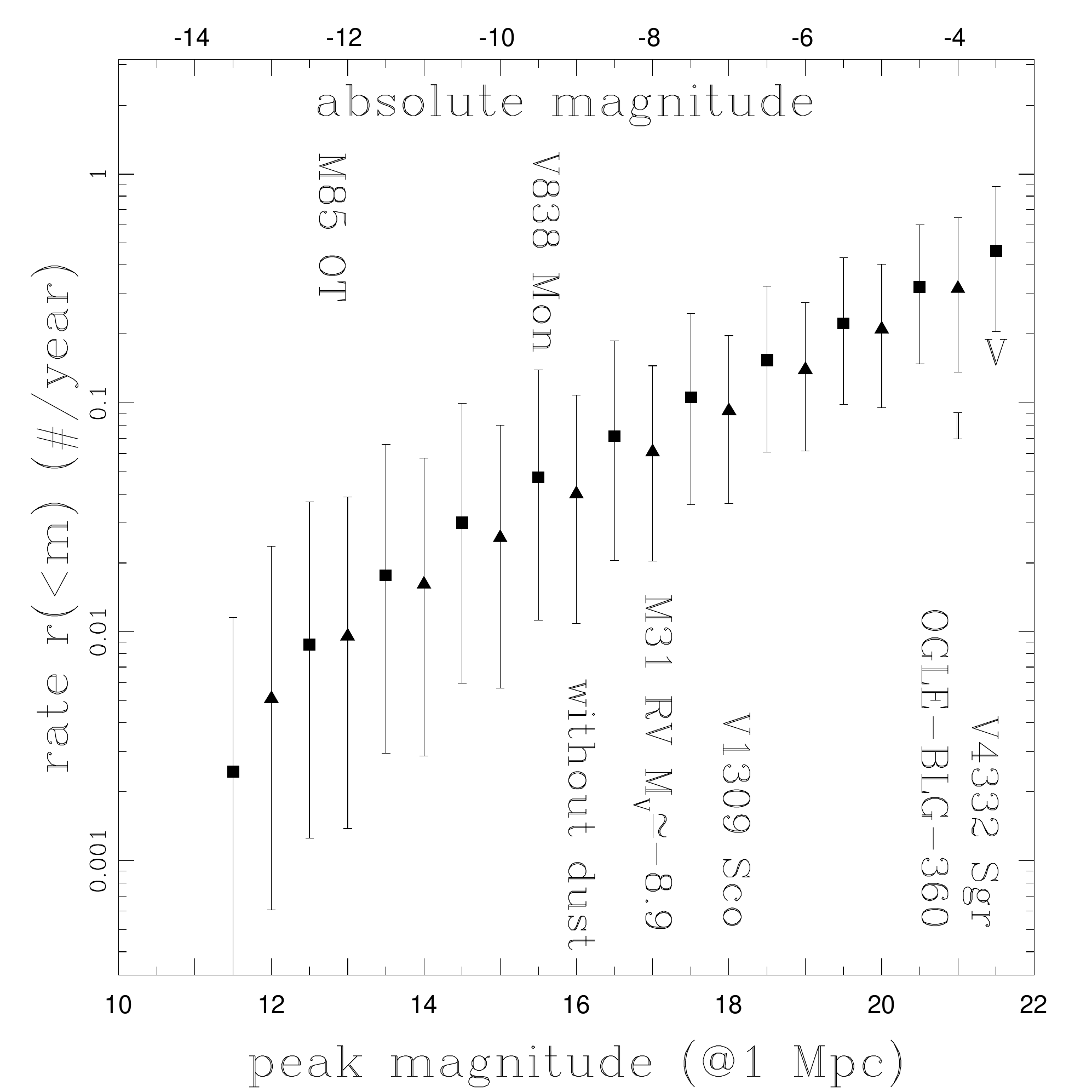}
\caption{
  Integral event rates $r(<m)$  as a function of peak absolute magnitude (upper
  scale) or at a distance of 1~Mpc (lower scale) roughly
  corresponding to M~31 or M~33.  The V band (squares)
  and I band (triangle) rates are shown on a staggered
  grid averaged over the thin disk $m_{lim}=13$~mag MCMC models.
  The error bars symmetrically encompass 90\% of the trials
  and are strongly correlated.  The apparent magnitudes of
  the M~31~RV are shown with and without correcting for the
  estimated foreground extinction. The Galactic events
  and the M85~OT
  are shown at their estimated V band absolute magnitudes.
  }
\label{fig:m31}
\end{figure*}

\begin{figure*}
\includegraphics[width=3.5in]{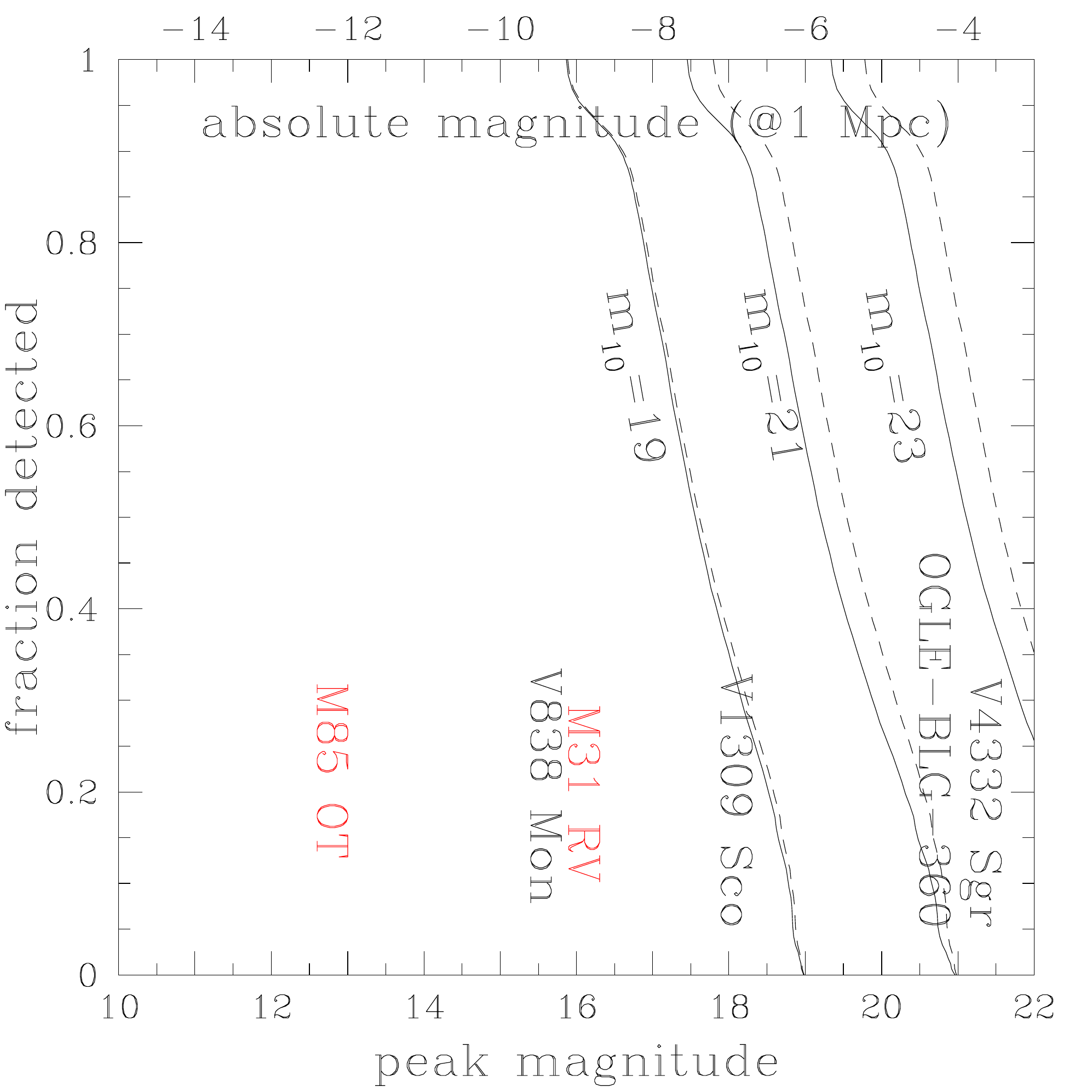}
\caption{
  Estimated completeness as a function of apparent magnitude (lower axis)
  or converted to the absolute magnitude corresponding to a distance 
  of 1~Mpc (upper axis) for a galaxy with the surface brightness profile 
  of M~31 and surveys with an empty field $S/N=10$ at $m_{10}=19$, $21$ or $23$~mag in 
  either V (solid) or I (dashed) combined with a $S/N=10$ detection threshold. 
  PTF has $m_{10} \simeq 20$~mag, the POINT-AGAPE variability survey
  of M~31 (\citealt{An2004}) had $m_{10}\simeq 24$~mag, and the
  LBT variability survey (\citealt{Kochanek2008}) has $m_{10}\simeq 26$~mag. 
  The labels for the various transients are placed at their peak
  V band absolute magnitude.  For a galaxy at a different distance,
  the absolute magnitude scale should be shifted by the appropriate
  factor (e.g. at a distance of 10~Mpc, $-10$ shifts to $-15$).
  }
\label{fig:m31comp}
\end{figure*}

\begin{figure*}
\includegraphics[width=3.5in]{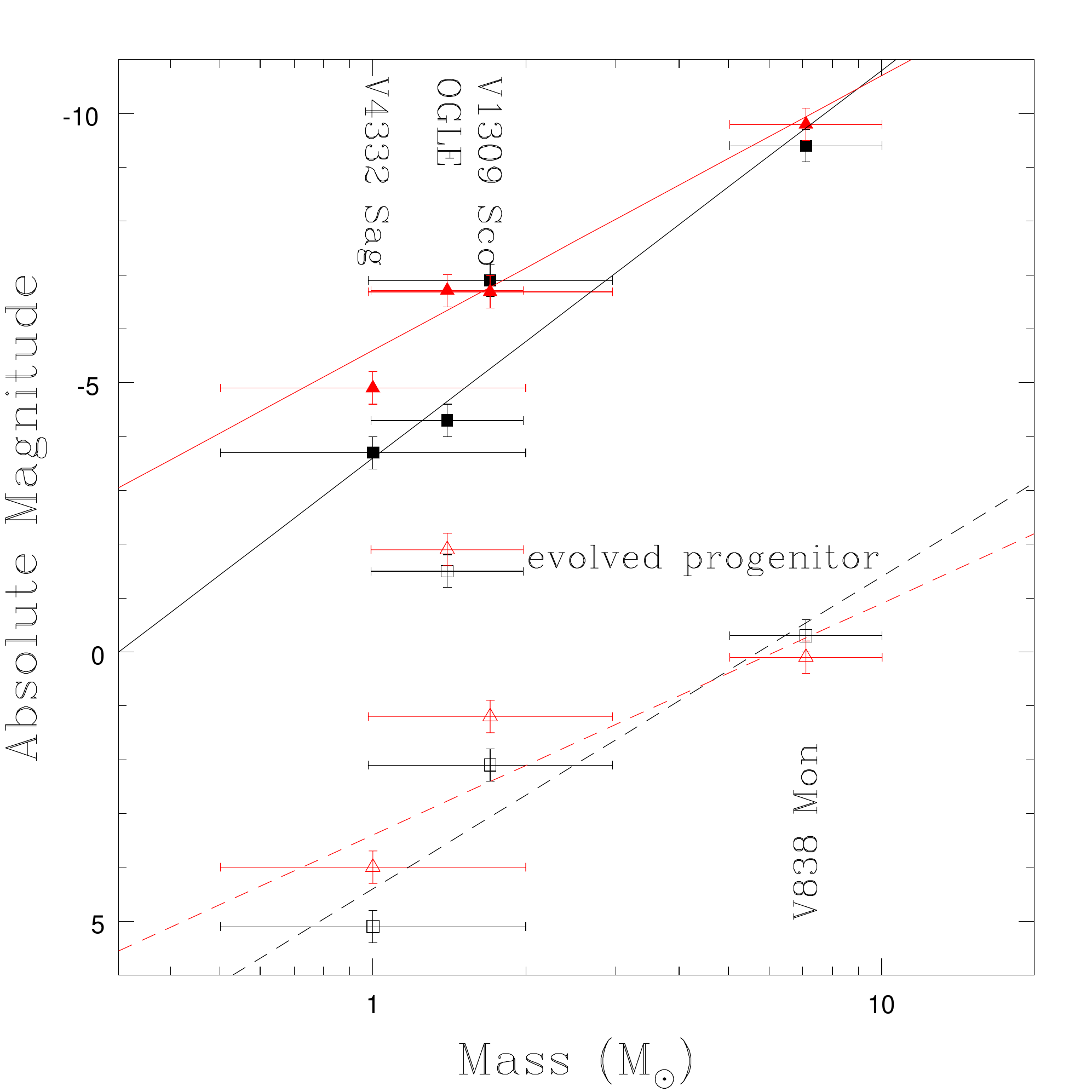}
\caption{
  Absolute magnitudes of the progenitors (open symbols) and transient peaks 
  (filled symbols) in the V (squares) and I (triangles) bands as a function
  of the progenitor mass estimates.  The best
  power law fits are also shown.  The progenitor fits exclude OGLE-2002-BLG-360
  because it must be an evolved star given its assumed distance.  The
  M85 OT2006-1 would not follow these trends since \protect\cite{Ofek2008} 
  estimate it must have $M<7 M_\odot$ while the transient peaked at
  $M_V \simeq -12$~mag.  
  }
\label{fig:lum}
\end{figure*}

\begin{figure*}
\includegraphics[width=3.5in]{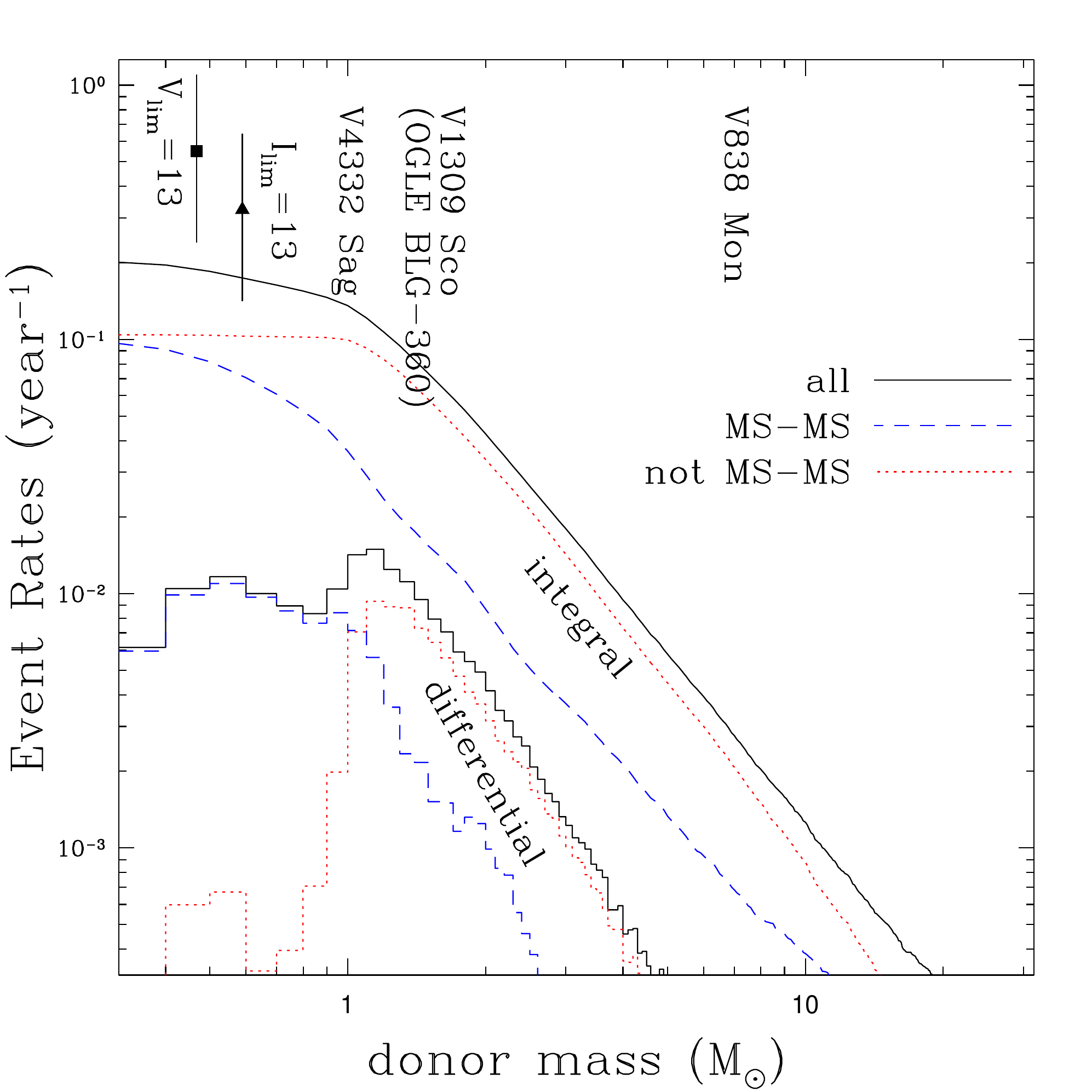}
\caption{
  Binary population synthesis models of the differential (lower curves)
  and integral (upper curves) Galactic event rates for all (black solid), 
  MS-MS (blue dashed) and not MS-MS (red dotted) binary progenitors.
  The filled square (filled triangle) shows the V band (I band) 
  observational estimates for the Galactic merger rates assuming
  the thin disk $m_{lim}=13$~mag models.  The labels
  for the four Galactic events are located at their nominal mass
  estimates.  Three of the observed events are MS-MS and one,
  OGLE~2002-BLG-360, shown in parenthesis, is a MS-evolved event.
  }
\label{fig:rates}
\end{figure*}

Figures~\ref{fig:prob} and \ref{fig:detect} illustrate the resulting
detection probabilities as a function of either apparent or absolute
magnitude for the thin disk spatial distribution.  Figure~\ref{fig:prob} shows that most of the candidates
have similar peak apparent magnitudes, and that the fraction of 
events that will be this bright is a small fraction of all events
with their absolute magnitudes.  The exception is the one source
found in a deep, modern survey, OGLE~2002-BLG-360.  As the name
indicates, however, the OGLE survey only covers a small region in
the Galactic bulge.  Note that we frequently abbreviate OGLE~2002-BLG-360 to 
OGLE-BLG-360 or just OGLE.  Because this transient
was found in a narrow, deep survey, we only compute the rates
using the other three Galactic transients.

Clearly the effective limiting magnitude of
the searches for these sources is a key parameter in the completeness
corrections.   Figure~\ref{fig:detect} illustrates this by showing
the fraction of sources detected at a given absolute magnitude as
a function of the limiting magnitude $m_{lim}$.  The completeness
for events like V838~Mon, the most luminous Galactic event, is 
far higher than for events like V4332~Sgr.  The absolute magnitude
distribution of the observed events already shows evidence that 
fainter events are more common, which means that the actual
luminosity function must be rising for fainter sources.  

With these assumptions, the expected number of events is
\begin{equation}
   N =  { A r \Delta t f_{sky} \ln 10 \over 2.5}
          \int_{M_1}^{M_0} d M \left( { L \over L_0} \right)^{1-x} f(M)
     \label{eqn:expect}
\end{equation}
where $M_1 < M < M_0$ is the absolute magnitude range corresponding
to $L_0 < L < L_1$, and the probability of observing an event of absolute magnitude
$M$ is
\begin{equation}
    P(M) = \left( { L \over L_0} \right)^{1-x} f(M) 
       \left[ \int_{M_1}^{M_0} d M \left( { L \over L_0} \right)^{1-x} f(M) \right]^{-1}.
      \label{eqn:pmag}
\end{equation}
Combining these, the probability of the observed data given the
model parameters $p$ is
\begin{equation}
    P(D|p) \propto  { N^3 \over 3! } e^{-N} \Pi_{i=1}^3 P(M_i)
\end{equation}
where the first term is the Poisson probability of three events
given the expectation value $N$ (Equation~\ref{eqn:expect})
and the second term is the product
of the probabilities of the observed absolute magnitudes of the three
events (Equation~\ref{eqn:pmag}).  The Bayesian probability of the parameters given the
data is then
\begin{equation}
    P(p|D) \propto P(D|p) P(p)
\end{equation}
where $P(p)$ describes the priors on the parameters for the rate ($r$)
and the luminosity function ($x$, $M_0$/$L_0$ and $M_1/L_1$).  We used a logarithmic
prior for the rate, $P(r) \propto 1/r$, and a uniform prior for $x$.
We simply fix $M_0$ and $M_1$ to roughly encompass the observed magnitude
ranges. If optimized, they would converge to values exactly bracketing 
the observed absolute magnitude range.  We optimize the model and estimate 
the uncertainties using Markov Chain Monte Carlo (MCMC) methods.  The
resulting estimates for the rates, completeness and slope of the
luminosity function are provided in Table~\ref{tab:results}.

We first fit the V band luminosities, fixing
the magnitude range to $M_0=-3$ and $M_1=-14$.  The
primary systematic uncertainty other than $f_{sky}$ is
the limiting magnitude.  As our standard model, we
will use the thin disk spatial distribution and 
$m_{lim}=13$~mag to roughly match the pre-transient 
detection limits discussed in the Appendix.  This
results in a model with $x \simeq 1.4 \pm 0.3$, a median 
rate of $r=0.53$~year$^{-1}$ ($0.22 < r < 1.1$) and a 
median completeness of $C=0.21$ ($0.15 < C < 0.30$) 
where we report 90\% confidence limits.  And, as 
noted earlier, the rate estimates can be rescaled
as $r \propto (25/\Delta t) f_{sky}^{-1}$.  

If anything, however, Figure~\ref{fig:prob}
seems to suggest that $m_{lim}$ should actually be lower
than $13$~mag, which is plausible given that pre-transient
detection limits will generally be an overestimate of the
magnitude at which a transient will actually be identified.
If we reduce the magnitude limit to $m_{lim} =11$ or $9$~mag,
the median rates increase to $0.99$~year$^{-1}$ ($0.36 < r < 2.4$)
or $2.3$~year$^{-1}$ ($0.67 < r < 7.0$) and the
median completenesses decrease to  $C=0.11$ ($0.07 < C < 0.19$) 
and $C=0.04$ ($0.02 < C < 0.12$), respectively.  The
slope becomes somewhat shallower as we increase the magnitude
limit, with
$x=1.6\pm0.5$, $1.5\pm0.3$ and $1.4\pm0.3$ for $m_{lim}=9$,
$11$ and $13$~mag.  There is also a covariance between 
$r$ and $x$ for fixed $m_{lim}$ that steeper slopes allow higher total rates 
because more systems are intrinsically faint.

We obtain similar results if we use the I band magnitudes
after shifting the absolute magnitude range by the typical
V$-$I color to $-4 < M_I < -15$.  For limiting I band
magnitudes of $m_{lim}=9$, $11$ and $13$~mag we find
$r=1.9$ ($0.47 < r < 6.5$),
$r=0.66$ ($0.27 < r < 1.4$), and
$r=0.32$ ($0.14 < r < 0.63$)~year$^{-1}$, respectively.
The slopes of $x=1.7\pm0.5$, $1.5\pm 0.3$ and $1.4\pm 0.3$ 
slowly become shallower with increasing $m_{lim}$ but
are all consistent with a slope of $x \simeq 3/2$. The
median completeness rises from $C=0.045$ and $0.17$ to $0.35$,
moderately higher than the V band estimates.    

Table~\ref{tab:results} also provides the results using
the thick$+$thin disk model of the source distribution.  Since the
little extincted thick disk population provides half
of the event rate, the completenesses are now much higher,
and the required rates are lower by factors of 2--3. 
As discussed earlier, we view the thin disk model as
more representative of the observed distribution of 
sources and certainly of all the higher mass progenitors.
Unfortunately, the likelihoods provide no useful 
discrimination between the models.

We adopt the thin disk model with $m_{lim}=13$~mag as
our fiducial case largely because it lies roughly in
the middle of the various estimates in Table~\ref{tab:results}. 
Thus, the rates for $M_V/M_I<-3$/$-4$~mag are $0.5/0.3$~year$^{-1}$
with statistical uncertainties that are basically
Poisson and systematic uncertainties of roughly a factor
of two.  Events comparable to or brighter than
V1309~Sco have a rate of $r=0.13$~year$^{-1}$ 
($0.047 < r < 0.28$ for $M_V=-7$~mag) or
$r=0.061$ ($0.020< r < 0.15$ for $M_I=-8$~mag).
Events comparable to or brighter than V838~Mon
have a rate of $r=0.034$~year$^{-1}$
($0.0082 < r < 0.12$ for $M_V=-10$~mag) or
$r=0.016$ ($0.0028< r < 0.057$ for $M_I=-8$~mag).
Roughly speaking, a stellar merger occurs every few
years, there is one as luminous as V1309~Sco once per
decade and there is one as luminous as V838~Mon every
forty years.  These integral rate estimates are shown
in Figure~\ref{fig:m31} and they can be converted to
magnitude limited rates in the Galaxy using 
Figure~\ref{fig:detect}.

Figure~\ref{fig:m31} also shows the apparent magnitudes
the transients would have at a distance of 1~Mpc,
roughly matching the distances to M~31 or M~33.
The V band properties of the M~31~RV and the
M~85~OT are shown for comparison.  No extinction
is included in these rate estimates.  The apparent
magnitudes must
be shifted to include any applicable foreground 
extinction from either our Galaxy or any host galaxy. 
Assuming M~31 is a twin of our Galaxy, the rate 
of events like the M~31~RV is intermediate to events
like V1309~Sco and V838~Mon, so roughly one every
20 years.  In theory, these nearby galaxies 
should be ideal laboratories for better estimating
merger rates because the extinction corrections 
become smaller and simpler.  

We can estimate the completeness of searches in an external
galaxy as follows.  We adopt the r-band major axis surface
brightness profile of M~31 from \cite{Kent1987} and convert
it to an estimated V and I band profile as $V=r+0.41$ and
$I=r-0.96$ based in the color conversions in \cite{Fukugita1995}.
For a population distributed like the r-band surface brightness,
we can simply integrate the profile to determine the fraction
of the luminosity in regions above a given surface brightness.  For a survey
with a detection limit $m_{10}$ corresponding to a signal-to-noise ratio of
$S/N=10$ in a empty field, the limiting magnitude for a
signal-to-noise ratio of $S/N$ including the surface 
brightness of the galaxy is
\begin{equation}
    m = m_{10} -2.5 \log_{10} G
\end{equation}
where
\begin{equation}
    G = 
  { \hat{N}^2 + \hat{N} \left(\hat{N}^2 + 4 \hat{s}' + 4\hat{s}\hat{s}'\right)^{1/2},
     \over 2 \left(1+\hat{s}\right) },
\end{equation}
$\hat{N}=SNR/10$, $\hat{s}=A 10^{-0.4(\mu_s-m_{10})}$ is the
ratio of the sky flux to the source flux, 
$\hat{s}' = \hat{s} + A 10^{-0.4(\mu_g-m_{10})}$ 
is the ratio of the sky plus galaxy flux to the source flux,
and $ A = \pi FWHM^2$ is the area of the photometry aperture.
We adopt sky brightnesses of $21.8$~V~mag/arcsec$^2$ and
$19.5$~I~mag/arcsec$^2$ and set $FHWM=1\farcs5$.  Figure~\ref{fig:m31comp}
shows the results for surveys with $SNR=10$ at $m_{10}=19$,
$21$ and $23$~mag where we also require $SNR=10$ 
for a detection (so $\hat{N}=1$).  

Figure~\ref{fig:m31comp} shows that it is relatively easy to
detect events like V838~Mon or the M31~RV in nearby galaxies
and the completeness for events like V1309~Sco will be 
reasonably high.  PTF, for example, 
tries to survey M~31 on a nightly basis, with an 
empty field $S/N=5$
at $21$~mag (\citealt{Law2010}), which would correspond 
to $m_{10} \simeq 20$~mag if we factor in its poorer image 
quality.  At this depth it is easy to detect the luminous
but rarer events like V838~Mon or the M~31~RV, possible
to detect events like V1309~Sco, but far to shallow to
detect the common events like V4332~Sgr.  This is further 
complicated by the large numbers of other variable
stars with similar luminosities to these faint transients
(see, e.g., the M~31 variability survey by \citealt{An2004}).  

For more distant galaxies, we simply shift the absolute
magnitude scale at the top of Figure~\ref{fig:m31comp}
to the right.  So if a survey like PTF begins to have severe completeness
problems at $M_V \simeq -7$~mag at a distance of 1~Mpc,
the limit is roughly $M_V \simeq -9$~mag at 3~Mpc (e.g., M81)
and $M_V\simeq -12$~mag at 10~Mpc.  Substantially 
deeper surveys are needed to detect the more common
but fainter events.  Nonetheless, for a luminosity
function declining as $L^{-x}$, the number of events
out to distance $d$ grows as $d^{5-2x} \sim d^2$ 
for our estimate of $x \simeq 3/2$, giving a survey
like PTF considerable sensitivity to the bright
end of the luminosity function (objects like 
the M~85-OT, \citealt{Kulkarni2007}).

\section{Scalings With Mass }
\label{sec:properties}

The four Galactic objects also have progenitor mass estimates, and
Figure~\ref{fig:lum} shows the absolute magnitudes of the progenitors
and the transient peaks as a function of mass.  We assigned a mass 
corresponding to the geometric mean of the
estimates discussed in the Appendix and use the logarithmic spread of the
masses around this central value or a factor of two as the nominal
error on the mass. We simply assign $0.3$~mag as a nominal error
in the magnitudes, as this will be dominated by systematic uncertainties
in distances and extinction. Despite these various limitations, there
appear to be broad correlations of these magnitudes with mass.  For
the progenitors, they must exist because the mass estimates are derived
from the inferred luminosities.  OGLE~2002-BLG-360 stands out 
because it must be an evolved star given the assumed distance,
and because the rapidly changing color and multiple peaks of the 
transient lead to the estimated V and I band absolute magnitude 
peaks coming from different epochs (see \citealt{Tylenda2013}).

That there appear to be correlations suggest doing simple fits, and
we find that the peak magnitudes are well fit by 
\begin{eqnarray}
   M_V &= &(-3.6 \pm 0.8) - (7.2 \pm 1.5) \log\left( M/M_\odot \right)~\hbox{mag} \quad\hbox{and} \nonumber \\
   M_I &= &(-5.6 \pm 0.4) - (5.1 \pm 0.7) \log\left( M/M_\odot \right)~\hbox{mag} 
   \label{eqn:slope}
\end{eqnarray}
corresponding to $L\propto M^{2.9\pm0.6}$ and $\propto M^{2.0\pm0.3}$, respectively.
Here we have rescaled the (somewhat arbitrary) uncertainties in the mass estimates 
downwards by a factor of $0.7$ and $0.4$ so that the fits have a $\chi^2$ per degree 
of freedom close to unity.  This has little effect on the central values but makes
the parameter uncertainties consistent with the observed scatter.  Changing the
(also somewhat arbitrary) magnitude error has little effect.  For comparison,
the progenitors (excluding OGLE~2002-BLG-360 as an evolved star) are fit by 
\begin{eqnarray}
   M_V &= &(4.4 \pm 1.1) - (5.8 \pm 1.6) \log\left( M/M_\odot \right)~\hbox{mag} \quad\hbox{and} \nonumber \\
   M_I &= &(3.4 \pm 0.4) - (4.3 \pm 1.7) \log\left( M/M_\odot \right)~\hbox{mag},
\end{eqnarray}
although the I band fits are poor.   These fits are also shown in
Figure~\ref{fig:lum}.  The changes in magnitude are far weaker functions 
of the estimated masses, and to zeroth order the transient peaks are 8.5-9.0~mag 
brighter than the un-evolved progenitors almost independent of the mass.  Based
on the one example, OGLE~2002-BLG-360, the jump in luminosity for an evolved
star will typically be smaller. 
If M85~OT2006-1 is a member of this class, it
does not follow these correlations since \cite{Ofek2008} estimate that the
absence of a progenitor in archival HST images requires $M<7 M_\odot$ while
the transient peaked at $M_V\simeq -12$~mag.  This would be consistent with
the argument by \cite{Pastorello2007} that this transient was actually a
faint supernova or by \cite{Thompson2009} that it might be an example of
the SN~2008S class of transients.  

If we combine the luminosity function of the transients
$dN/dL \propto L^{-x}$ with the scaling of the transient
luminosity with mass, $L \propto M^a$, then the mass
function of the progenitors is $dN/dM \propto M^{-1+a(1-x)}$.
For our nominal values of $x \simeq 1.4\pm0.3$ and 
$a \simeq 2.5 \pm 0.5$ (where we are just splitting 
the difference between the V and I band slopes from 
Equation~\ref{eqn:slope}), this implies a mass function
of $dN/dM \propto M^{-2.0 \pm 0.8}$, consistent with 
typical IMFs.  Turning this around, matching the slope of a 
Salpeter IMF, $dN/dM\propto M^{-2.35}$, corresponds to 
$x=1.68$ or $1.45$ for $a=2$ or $3$. 

\section{Binary Population Synthesis Models}
\label{sec:theory}

Merger rates are implicit in any binary population synthesis model,
but appear never to have been explicitly presented.  Here we use
the {\tt StarTrack} population synthesis code \citep{Belczynski2002,Belczynski2008}
using the parameter choices described in \cite{Dominik2012}.
This model employs energy balance for CE evolution with a physical estimate for the 
donor's binding energy, updated wind mass loss prescriptions and a realistic mass 
spectrum for compact objects.  We make the very specific assumption that all
donors off the main sequence are allowed to survive CE (\citealt{Belczynski2007} 
and \citealt{Belczynski2010} discuss some alternative scenarios for Hertzsprung gap donors).
The IMF is a 3 component broken power law with
boundaries at $M_{\rm ZAMS} = 0.08$, $0.5$, $1.0$, and $150 M_\odot$ and slopes
of $-1.3$, $-2.2$ and $-2.7$ for the three mass ranges \citep{Kroupa2003}
and all results are for Solar metallicity ($Z=Z_\odot=0.02$).

We assume a binary fraction of 50\%, so that $2/3$ of all stars are in binaries.
We use a flat mass ratio distribution, $P(q)$ constant, with $0< q =M_2/M_1 < 1$
(e.g. \citealt{Kobulnicky2007}), a logarithmic distribution of binary separations
$P(a) \propto 1/a$ (e.g. \citealt{Abt1983}) ranging from where the primary just fills
its Roche lobe up to $10^5 R_\odot$, and a thermal-equilibrium distribution
of eccentricities $\Xi(e) = 2e $ for $0 < e < 1$ (\citealt{Duquennoy1991}).
We then evolve primaries with masses from $0.3$--$150M_\odot$ and secondaries
with masses from $0.08$--$150M_\odot$ assuming a constant star formation
rate of $3.5 M_\odot$~year$^{-1}$ for a period of $10$~Gyr.  This implies
a core collapse supernova rate of $0.018$~year$^{-1}$ that is compatible with other
estimates (see \citealt{Adams2013}).

We classify the stars as main sequence (MS) stars, evolved stars (off the main sequence
but retaining a hydrogen envelope), helium stars (no hydrogen envelope) and
compact objects (white dwarf, neutron star or black hole).  They are
referred to as the donor and companion stars following the standard 
terminology in binary evolution models.  For comparison to the observational
mass estimates we used the mass of the donor star, as it is almost always the
more massive, unless the system is a nearly equal mass ($q>0.8$) MS/MS
binary, in which case we used the total mass. Because
stellar luminosities are such strong functions of mass, the more
massive star always dominates the luminosity except for the
case of nearly equal mass MS/MS binaries.

Figures~\ref{fig:rates} shows the resulting Galactic rate estimates
in both integral and differential forms
for the overall population and divided into 
MS/MS binaries and systems with at least one evolved component
(``not MS/MS'').  Table~\ref{tab:rates} breaks
the rates down in more detail and by whether the stars ultimately
merge.  The total rate in this model is $0.2$~year$^{-1}$, which is
compatible with our observational estimates from \S2.  Of these, 70\% result
in a final merger and 30\% leave a binary.  Events with two main sequence
stars (44\%) or a main sequence star and an evolved star (42\%)
dominate the rates.  Most of the remaining events are between compact objects
and main sequence (5\%) or evolved (9\%) stars.  Events with
helium stars should be very rare. 

Interactions with evolved stars dominate at higher masses -- high
mass MS-MS
events like V838~Mon should be the minority (20\%).  Given the age
of the Galaxy, main sequence events have to dominate the rates at
low masses because the stars have not had time to undergo nuclear
evolution.  This does not mean that there are no mergers of low mass stars because
strong magnetic breaking associated with the deep convective envelopes
of low mass MS stars (\citealt{Ivanova2003}), gravitational radiation
and tidal interactions can still lead
to orbital decay and a CE event.  The predicted rates peak for
stars near $1$-$2M_\odot$, just where we find V4332~Sag, OGLE~2002-BLG-360
and V1309~Sco.  Above $M\sim M_\odot$ the rate declines slightly
steeper than the input IMF, roughly as $M^{-3}$ instead of $M^{-2.7}$.
This is also broadly consistent with our inferences from the 
observed events in \S\ref{sec:properties}.  The observed events
are dominated by MS-MS events (three of four), but this is only
mildly unlikely given the prediction that there are roughly equal
numbers of each type.  In a more detailed model of the spatial
distribution, the MS-MS events are likely to be more visible
because they will have higher average scale heights and less
extinction than MS-evolved star mergers.

\section{Discussion}
\label{sec:discuss}

Much of binary evolution depends on the rate at which stars merge or enter into CE
phases.  It appears likely that a significant fraction of stars must to so, in which case
the rate of stellar mergers in the Galaxy must be quite high.  If the class of
objects encompassing V4332~Sgr, V838~Mon, OGLE~2002-BLG-360 and V1309~Sco are
examples of such events, as seems almost certain following the direct observations
of the merger of V1309~Sco (\citealt{Tylenda2011}), then the observed rate of 
such mergers is also high.  Using the model we developed in \cite{Adams2013} to
estimate the visibility of supernovae in the galaxy, we can correct the observed
rate for completeness to find that the rate of mergers brighter than $M_V=-3$~mag
($M_I=-4$~mag) is roughly one every 2--3 years, albeit with significant uncertainties
associated with the definition of the ``survey'' conditions under which they were
found.  That all the transients found outside the OGLE survey are very bright
($m_I<7$~mag) at peak, while the one found by the OGLE bulge survey is over an order
of magnitude fainter, confirms that there must be many fainter events elsewhere
in the Galaxy.

The luminosity function of the transients is roughly $dN/dL \propto L^{-3/2}$,
so there are intrinsically many more low luminosity events like V4432~Sgr than
high luminosity events like V838~Mon or V1309~Sco.  The rate of these bright
events is roughly one every 10--50 years, which roughly corresponds to crude 
rate estimates by \cite{Soker2006} and \cite{Ofek2008}.  As emphasized by 
\cite{Kulkarni2007}, the distinctive post-peak evolution of these transients
to being extremely cold M/L class supergiants combined with dust formation 
provides a good means of recognizing these events.   Some care will be required
because the SN~2008S class of transients are also lower luminosity than typical
supernovae and form dust (see \citealt{Prieto2008}, \citealt{Thompson2009}, \citealt{Kochanek2011}).
While these transients become very red, it appears to be solely due to dust
absorption - a cold stellar photosphere has never been observed.

In the Galaxy, finding more examples is simply limited by the lack of complete,
deeper surveys of the Galactic plane.  I band (or even V band) surveys of the
Galaxy that were complete even to $m_{lim}\simeq 16$~mag would represent a significant
improvement over the depth of the searches implied by the existing events
(see Figure~\ref{fig:detect}). 
Considerable progress would be made if ASAS-SN (\citealt{Shappee2014}),
ATLAS (\citealt{Tonry2011}) or PTF (\citealt{Rau2009}, \citealt{Law2010}) systematically surveyed
large fractions of the Galactic plane.  LSST, unfortunately, will not
emphasize variability surveys of the Galactic plane (see the critique
by \citealt{Gould2013}).  At the moment, the best survey for Galactic
events is probably being carried out by Gaia (e.g., \citealt{Eyer2011}).  
A great advantage of many of the Galactic events
is that it will be possible to characterize the progenitor stars. 

There is no simple way to characterize existing
variability surveys of nearby galaxies like M~31.  In a simple model for 
the completeness (see Figure~\ref{fig:m31comp}), a survey like PTF with
$m_{10} \simeq 20$~mag should be able to find events like 
V838~Mon, the M~31~RV transient and V1309~Sco, but not significantly fainter
transients like OGLE~2002-BLG-360 or V4332~Sgr.   Thus, while the overall rate
in M~31 is probably comparable to that of the Galaxy, the rate of mergers
detectable by PTF is roughly one per decade.  By surveying many galaxies,
PTF can have a significant rate of more luminous events (e.g. \citealt{Rau2009})
but this assumes that the luminosity function extends to $M_V \sim -12$~mag
or brighter.

Deeper variability surveys like the POINT-AGAPE microlensing survey
of M~31 ($m_{10}\simeq 24$~mag, \citealt{An2004}) or our variability
survey of nearby galaxies with the Large Binocular Telescope
($m_{10}\simeq 26$~mag, \citealt{Kochanek2008}) can find significantly
fainter events.  The difficulty for these faint events will be that
crowding in an external galaxy will make it difficult to separate
merger events from other red variable sources.  In the Galaxy this
is relatively straight forward because no sources other than novae show such
dramatic increases in flux, but this depends on
having a large dynamic range.  Archival HST photometry may supply
this in some cases, and the peculiar properties of the transients
may be sufficient to distinguish mergers from other sources of 
variability (\citealt{Kulkarni2007}).  In any case, if these rate
estimates are correct, our LBT survey must already contain several
merger events and their remnants.

Even with so few examples, we can already identify several interesting 
correlations.  First, the luminosity function of
the transients is roughly $dN/dL \propto L^{-1.4\pm0.3}$.  Second, the
peak luminosities increase rapidly with the progenitor mass, with 
$L \propto M^{2-3}$.  This means that the transient peaks roughly 
track the main sequence luminosity of the progenitors but are
$2000$-$4000$ times brighter.  Essentially, the photosphere seems to
expand by a huge factor ($\sim 50$-$100$) but the photospheric 
temperature cannot drop by a huge factor, so to zeroth order the
transient peak is simply the main sequence luminosity multiplied
by a large number.   Third, the mass function of the merger
progenitors is $dN/dM \propto M^{-2.0 \pm 0.8}$, consistent
with typical IMFs.

When we compare these to {\tt StarTrack} binary population 
synthesis models \citep{Belczynski2002,Belczynski2008} simply
using the parameter choices from \cite{Dominik2012}, we find
remarkably good agreement between the observations and the
predictions.  In particular, the predicted rate of $0.2$~year$^{-1}$
is high and the progenitor mass function declines roughly
like the IMF.  The models also predict that the rates peak
at roughly $M \sim M_\odot$ with higher mass events dominated
by systems that include at least one evolved star and lower
mass events dominated by mergers of main sequence stars.  
The rates are dominated by MS-MS (44\%), MS-evolved (42\%),
evolved-compact (9\%) and MS-compact (5\%) events with
negligible contributions from other possibilities.  Of those
entering a CE phase, 70\% are predicted to merge.  

There are many uncertainties in these theoretical estimates.
Some of the major ones are uncertainties in the initial
orbital distributions, star formation histories,
stellar models and the expansion
rates of stars, magnetic braking and tidal interaction
models, and the development of dynamical instabilities
during Roche lobe overflow leading into a CE event.  For
example, the orbital period distributions found by
\cite{Sana2012} would have more close binaries and hence
more CE events for high mass stars than in our models.
Overall, roughly 1 in 5 binaries in our models go through
a CE interaction, corresponding to 2/15 of all stars
and 1/10 of stellar systems.  If we allowed {\it all} stars to be
in binaries and {\it all} stars to undergo a CE event,
we get a maximum possible Galactic rate of $\sim 1$-$2$
CE events per year that could not be exceeded without
carefully tuning the Galactic star formation history.     
Future observations to better measure the rates of these events,
their progenitor mass functions and the evolutionary
states of the progenitors will be a powerful constraint
on many of these uncertainties.

\section*{Acknowledgments}
We would like to thank N. Ivanova as the organizer of the conference ``Stellar Tango in the
Rockies" and conversations with M. Cantiello and J. Staff at the conference.
We also had valuable discussions with A.~Gould, E.~Ofek, O.~Pejcha,
M.~Pinsonneault, K.Z.~Stanek, and T. Thompson.
KB acknowledges support from a Polish Science Foundation ``Master2013''
Subsidy, Polish NCN grant SONATA BIS 2, NASA Grant Number NNX09AV06A 
and NSF Grant Number HRD 1242090 awarded to the Center for Gravitational 
Wave Astronomy at U.T.~Brownsville.

\appendix
\section{Summary of Objects}

Here we provide short summaries of the properties of the objects as reported in Table~\ref{tab:summary}.
We start with the Galactic transients ordered by date and then the two extragalactic
candidates.  Most of these events were found by amateurs (V3332~Sgr, V838~Mon and
V1309~Sco) or serendipitously (M31~RV).  OGLE-2002-BLG-360 was found as part of the
OGLE microlensing survey (see \cite{Tylenda2013} for its history) and the M~85  
transient was found a part of the PTF survey, which is partly motivated by 
searching for stellar mergers (\citealt{Rau2009}).

\begin{itemize}
\item V4332~Sgr (February 1994) was discovered by \cite{Hayashi1994} 
at $V\simeq 8.4$~mag with no estimate of prior magnitude limits.  The transient
peaked at $V \simeq 8.5$ and $I\simeq 6.9$~mag and has an estimated foreground
extinction of $E(B-V)\simeq 0.3$~mag (\citealt{Martini1999}).  The distance
is uncertain, but \cite{Tylenda2005a} argue for $d \simeq 1.8$~kpc where it
would be a G/K main sequence disk star, implying masses $\sim 1M_\odot$ with
$V\simeq 17.3$~mag and $I\simeq 15.8$~mag.

\item V838~Mon (January 2002) was discovered by \cite{Brown2002} at a (photographic) magnitude of $\sim 10$~mag
  with a limiting depth for the closest prior observation of $\sim 12$~mag.  It reached peak magnitudes
  of $V \simeq 7.0$~mag and $I_c \simeq 5.5$~mag (\citealt{Munari2002}).  Light echoes (\citealt{Bond2003})
  enabled an accurate determination of the distance to be $d \simeq 6.1 \pm 0.6$~kpc (\citealt{Sparks2008}).
  Several lines of evidence lead to an estimated foreground extinction of $E(B-V)\sim 0.7$ to $0.9$~mag
  (e.g. \citealt{Munari2005}, \citealt{Tylenda2005b}).  The more luminous component of the merger was
  probably a $5$-$10 M_\odot$ B star (\citealt{Tylenda2005b}) and there is no direct constraint on the
  secondary.  From this study we adopt $I=15.8$ and estimate $V=17.3$~mag.
  As expected for such a young star, V838~Mon lies in the Galactic plane.

\item OGLE-2002-BLG-360 (October 2002) was initially considered a long duration microlensing
event and then later realized to be a likely stellar merger (\citealt{Tylenda2013}).  The
OGLE trigger occurred at $I\simeq 15.5$~mag, as the object brightened from a baseline at $I\simeq 16$~mag,
and then peaked at $V \sim 16.5$ and $I \sim 11.3$~mag, making it much redder near peak than
most of the other merger candidates. In fact, the V band and I band peaks occur at
different epochs. \cite{Tylenda2011} adopt a foreground extinction of 
$E(B-V) \simeq 1$~mag and place the event in the Galactic bulge at $d=8.2$~kpc in the absence
of any direct constraints.  They also found that the progenitor showed a significant mid-IR
excess which, combined with its red color, implied the presence of $\tau_V \sim 3$ of circumstellar
dust in addition to the foreground component.  We treat this source as if it has a total 
extinction of $E(B-V) \simeq 2$~mag.  \cite{Tylenda2013} model the progenitor as
a $L \simeq 300L_\odot$ K giant with $V=19.3$~mag and $I=16.1$~mag,
 which would correspond to a relatively low mass $1$-$2M_\odot$
evolved star.  

\item V1309~Scorpii (September 2008) was independently discovered by Nishiyama \& Kabashima at (unfiltered) 9.5~mag
  with prior upper limits of 12.8~mag and by Sun et al. at 10.5~mag with prior upper limits of 13.5~mag
 (\citealt{Nakano2008}).  The event peaked at $V\simeq 8$~mag and $I\simeq 7$~mag (see \citealt{Mason2010},
  \citealt{Tylenda2011}).  \cite{Mason2010} pointed out that the properties of the transient were very
  similar to those of V838~Mon.  In this case, pre-explosion light curves clearly show that the 
  transient was caused by a stellar merger, probably two K giants with a total mass of $1$-$3M_\odot$
  at a distance of $3.0\pm0.7$~kpc with $V=17.0$~mag and $I=14.9$~mag (\citealt{Tylenda2011}). This mass range seems to be consistent
  with theoretical examinations of the binary evolution (\citealt{Stepien2011}, \citealt{Nandez2013},
  \citealt{Pejcha2013}).  Extinction estimates range from 
  $0.6 \ltorder E(B-V) \ltorder 1.0$~mag (\citealt{Mason2010}, \citealt{Tylenda2011}).  Given the
  estimated distance and its Galactic coordinates, this event is also associated with the Galactic
  disk.  

\item M31~RV (``Red Variable'') was found by \cite{Rich1989} in September 1988 as a new source 
  with $i\sim 14.9$ and $g \sim 16.9$~mag.   With additional data by \cite{Mould1990},
  \cite{Bryan1992} and \cite{Tomaney1992}, the transient peak was near $R \simeq 16$
  and $B \simeq 18$~mag followed by dust formation and a rapid fading. 
  \cite{Rich1989} estimated $E(B-V) \simeq 0.3$~mag and the distance is 
  $d \simeq 1$~Mpc.  Using
  $T_e \simeq 4500$~K, typical of these transients at peak, we converted these
  peak magnitudes to $V \simeq 17.0$ and $I \simeq 15.5$~mag. 

\item M85~OT2006-1 (January 2006) peaked at $R \simeq 19.0$ and $I \simeq 18.5$~mag (\citealt{Kulkarni2007})
  with and estimated foreground extinction of $E(B-V) \simeq 0.1$-$0.2$.  Based on the absence
  of the progenitor in archival Hubble Space Telescope data, \cite{Ofek2008} argue that the
  progenitor must have $M < 7M_\odot$ and suggest $ M \sim 2 M_\odot$.  Again using 
  $T_e \simeq 4500$~K, we converted the $R$ magnitude to $V \simeq 19.7$~mag. 

\end{itemize}

\vfill\eject

\onecolumn
\begin{deluxetable}{lrrrcrrrrc}
\scriptsize
\tablecaption{Properties of Merger Candidates}
\tablewidth{0pt}
\tablehead{
  \multicolumn{1}{c}{Object} &
  \multicolumn{2}{c}{Peak Mag} &
  \multicolumn{1}{c}{dist} &
  \multicolumn{1}{c}{$E(B-V)$} &
  \multicolumn{2}{c}{Peak Abs Mag} &
  \multicolumn{2}{c}{Progenitor} &
  \multicolumn{1}{c}{Mass} \\
  \multicolumn{1}{c}{} &
  \multicolumn{1}{c}{V} &
  \multicolumn{1}{c}{I} &
  \multicolumn{1}{c}{(kpc/Mpc)}  &
  \multicolumn{1}{c}{(mag)} &
  \multicolumn{1}{c}{$M_V$} &
  \multicolumn{1}{c}{$M_I$} &
  \multicolumn{1}{c}{$M_V$} &
  \multicolumn{1}{c}{$M_I$} &
  \multicolumn{1}{c}{$M_\odot$} 
  }
\startdata
           V4332~Sag &$ 8.5$ &$ 6.9$ &$ 1.8$ &$0.30$ &$-3.7$ &$-4.9$  &$ 5.1$ &$ 4.0$ &$1$ \\ 
            V838~Mon &$ 7.0$ &$ 5.5$ &$ 6.1$ &$0.80$ &$-9.4$ &$-9.8$  &$-0.3$ &$ 0.1$ &$5-10$ \\ 
       OGLE~-BLG-360 &$16.5$ &$11.3$ &$ 8.2$ &$2.00$ &$-4.3$ &$-6.7$  &$-1.5$ &$-1.9$ &$1-2$ \\ 
           V1309~Sco &$ 8.0$ &$ 7.0$ &$ 3.0$ &$0.80$ &$-6.9$ &$-6.7$  &$ 2.1$ &$ 1.2$ &$1-3$ \\ 
              M31~RV &$17.0$ &$15.5$ &$ 1.0$ &$0.30$ &$-8.9$ &$-10.0$ &       &       &$--$ \\ 
              M85~OT &$19.7$ &$18.5$ &$17.8$ &$0.20$ &$-12.2$ &$-13.1$&       &       &$<7$ \\ 

\enddata
\label{tab:summary}
\vspace{-0.25in}
\tablecomments{
   The Appendix provides a more detailed discussion, parameter ranges and references for
   each object.  The distances are in kpc (Mpc) for the Galactic (extragalactic) objects.
   We have included the estimated circumstellar extinction for OGLE-2002-BLG-360 in
   the estimate of $E(B-V)$.
   }
\end{deluxetable}

\end{document}